\documentclass[conference]{IEEEtran}
\IEEEoverridecommandlockouts

\usepackage{cite}
\usepackage{amsmath,amssymb,amsfonts}
\usepackage{algorithm}
\usepackage{algpseudocode}
\usepackage{graphicx}
\usepackage{textcomp}
\usepackage{booktabs}
\usepackage{xcolor}
\usepackage[bookmarks=false,colorlinks=true,citecolor=black,linkcolor=black,urlcolor=black]{hyperref}
\renewcommand{\url}[1]{\nolinkurl{#1}}
\renewcommand{\href}[2]{#2}
\usepackage{tikz}
\usepackage{pgfplots}

\pgfplotsset{compat=1.18}
\pgfplotsset{
  ieee single col/.style={
    width=8cm, height=7.5cm,
    tick align=outside,
    axis line style={line width=0.8pt},
    grid=both,
    major grid style={line width=0.4pt,draw=gray!50},
    minor grid style={line width=0.2pt,draw=gray!20},
    tick label style={font=\fontsize{8}{9}\selectfont},
    label style={font=\fontsize{8}{9}\selectfont},
    title style={font=\fontsize{8}{9}\selectfont},
    legend style={
      font=\fontsize{6}{7}\selectfont,
      at={(0.98,0.02)}, anchor=south east,
      draw=black, fill=white,
      fill opacity=0.85, text opacity=1,
      inner sep=2pt, row sep=-1pt,
    },
    legend cell align=left,
  }
}
\definecolor{colorTSDCEfft} {rgb}{0.00,0.27,0.67}
\definecolor{colorTSDCEdict}{rgb}{0.00,0.55,0.55}
\definecolor{colorOMP}      {rgb}{0.80,0.20,0.20}
\definecolor{colorMOMP}     {rgb}{0.95,0.55,0.15}
\usetikzlibrary{shapes.geometric, arrows.meta, positioning}

\AddToHookNext{shipout/foreground}{%
  \put(0.5\paperwidth,-0.24in){\makebox[0pt][c]{\parbox[t]{0.86\paperwidth}{\scriptsize\centering
    \copyright~2026 IEEE. Personal use of this material is permitted. Permission from IEEE must be obtained for all other uses, in any current or future media, including reprinting/republishing this material for advertising or promotional purposes, creating new collective works, for resale or redistribution to servers or lists, or reuse of any copyrighted component of this work in other works.\\
    Accepted for publication in the 2026 IEEE 1st Annual Integrated Sensing and Communication Conference (ISAC).}}}}

\begin{document}

\title{Low-Complexity Wideband Channel Estimation for ISAC in Analog Millimeter-Wave Systems

\thanks{Thanks to Grants PID2023-148671OB-I00 and PLEC2024-011160 funded by MICIU/AEI/10.13039/501100011033 and by ERDF/EU. This research was partially funded by the Generalitat Valenciana through the Prometeo project CIPROM/2026/037.}}

\author{
    \IEEEauthorblockN{Hamza Touati\IEEEauthorrefmark{1}, Raffaele Viterbo\IEEEauthorrefmark{2}, Mattia Brambilla\IEEEauthorrefmark{2}, Sandra Roger\IEEEauthorrefmark{1}, and Carmen Botella-Mascarell\IEEEauthorrefmark{1}}

    \IEEEauthorblockA{\IEEEauthorrefmark{1}Computer Science Department, Universitat de València, 46100 Burjassot, Spain}
    \IEEEauthorblockA{\IEEEauthorrefmark{2}Dipartimento di Elettronica, Informazione e Bioingegneria, Politecnico di Milano, 20133 Milan, Italy}

    \thanks{Corresponding author email: hamza.touati@uv.es.}
}

\maketitle
\begin{abstract}
Accurate wideband channel estimation is a key enabler for millimeter-wave (mmWave) integrated sensing and communication (ISAC) systems. In the narrowband scenario, the Transformed Spatial Domain Channel Estimation (TSDCE) method was previously shown to provide fast and accurate angle estimation with substantially lower complexity than conventional compressed-sensing approaches based on orthogonal matching pursuit (OMP). However, wideband ISAC scenarios additionally require delay estimation, making joint angle-delay sparse recovery computationally expensive due to the multidimensional dictionary search.
In this paper, we propose a wideband extension of TSDCE, namely TSDCE-TD, which incorporates time-domain support and decouples angle and delay estimation. Two variations are proposed for delay estimation: TSDCE-TD (Dict), which uses a one-dimensional delay dictionary, and TSDCE-TD (FFT), which applies Fast Fourier Transform (FFT) zero-padding interpolation. Path estimates are iteratively refined through successive interference cancellation and alternating angle-delay updates. Simulation results show that the proposed methods outperform OMP-based baselines under practical dictionary sizes, while achieving comparable performance at high-resolution settings with substantially lower computational complexity.  Moreover, the angle-delay estimates obtained using the proposed approach enable accurate single-anchor indoor localization in a ray-traced environment, achieving sub-20 cm positioning errors and satisfying practical indoor positioning requirements.
\end{abstract}


\section{Introduction}

Millimeter-wave~(mmWave) communication is fundamental to fifth-generation~(5G) and sixth-generation~(6G) wireless systems, leveraging the abundant spectrum in the 30--300~GHz band~\cite{hong2021role}. The short wavelengths enable compact antenna arrays with highly directional beamforming to overcome severe path loss propagation. Building on these capabilities, mmWave is increasingly explored for integrated sensing and communication~(ISAC)~\cite{Gon24}, where large bandwidths and directional beams provide fine spatial resolution, accurate target detection, and enhanced data rates~\cite{liu2025sensing}. A fundamental characteristic of mmWave propagation channels is their inherent sparsity, where only a few dominant propagation paths typically contribute to the received signal. This sparse structure has been widely exploited in channel estimation through compressed sensing and sparse recovery techniques, with orthogonal matching pursuit~(OMP) and its variants being among the most commonly adopted approaches. More recently, sparse recovery methods have also been extended to ISAC applications, including user localization~\cite{momp}, target detection~\cite{nokia2026}, classification~\cite{Barbieri_JC&S2026}, and environment sensing~\cite{tong2026distributed}. %
Despite their effectiveness, OMP-based approaches often suffer from relatively high computational complexity due to their iterative dictionary atom selection and signal reconstruction procedures, particularly in wideband large-scale mmWave systems. This motivates the development of lower-complexity sparse recovery techniques that can maintain accurate sensing and estimation performance while enabling faster processing for practical ISAC deployments \cite{momp}.

In this work, we present a low-complexity algorithm for sparse recovery in mmWave ISAC systems, referred to as TSDCE with Time-Domain support~(TSDCE-TD), which extends the Transformed Spatial Domain Channel Estimation (TSDCE) algorithm \cite{tsdce} to include delay estimation. Two variants are analyzed. First, TSDCE-TD (Dict) utilizes a one-dimensional delay dictionary, following an approach analogous to OMP-based methods but restricted solely to the delay dimension, where the computational cost scales only linearly with the grid size. Second, TSDCE-TD (FFT) leverages Fast Fourier Transform (FFT) zero-padding interpolation, thereby yielding the lowest computational complexity. The proposed approaches are evaluated against OMP-based baselines in terms of channel estimation accuracy, computational complexity, and user positioning performance within a ray-traced indoor scenario.
Simulation results show that TSDCE-TD achieves channel estimation accuracy comparable to or better than OMP-based methods, while reducing complexity by several orders of magnitude, and that the recovered parameters are sufficient for accurate indoor localization.

The remainder of the paper is organized as follows. Section~\ref{sec:system_model} introduces the wideband channel model and the analog beamforming signal model. Section~\ref{sec:algorithm} presents the proposed algorithms. Section~\ref{sec:perf_evaluation} provides simulation results and complexity analysis. Section~\ref{sec:conclusion} concludes the paper.

\section{System Model}
\label{sec:system_model}

We consider a single-user mmWave system in which the transmitter~(Tx) and receiver~(Rx) are equipped with Uniform Linear Arrays~(ULAs) of $n_{\mathrm{t}}$ and $n_{\mathrm{r}}$ antennas, respectively.
Due to the limited scattering characteristics of the mmWave propagation environment, the channel is composed of $L$ paths, each characterized by a complex gain $\alpha_l$, an Angle of Departure~(AoD) $\phi_l \in [0,2\pi]$, an Angle of Arrival~(AoA) $\psi_l \in [0,2\pi]$, and a propagation delay $\tau_l \geq 0$ measured in sampling-period units~\cite{venugopal2017channel}.
The wideband channel is frequency-selective, and its $d$-th delay tap for $d = 0,\ldots,D-1$ is~\cite{COMST2025}:
\begin{equation}
  \begin{split}
    \mathbf{H}[d] & = \sqrt{n_{\mathrm{t}} n_{\mathrm{r}}}
      \sum_{l=1}^{L}
      \alpha_l\,
      \mathbf{a}_{\mathrm{r}}(\psi_l)\,
      \mathbf{a}_{\mathrm{t}}^{\mathsf{H}}(\phi_l)\,
      p(d - \tau_l),
  \end{split}
  \label{eq:channel_td}
\end{equation}
where $p(\cdot)$ denotes the band-limited pulse-shaping filter response and $D$ is the maximum channel temporal support. The ULA steering vectors at Tx and Rx, assuming half-wavelength antenna separation, are defined as~\cite{tsdce}:
\begin{align}
  \mathbf{a}_{\mathrm{t}}(\phi_l)
  &= \frac{1}{\sqrt{n_{\mathrm{t}}}}
     \bigl[1,\,e^{-j\pi\cos\phi_l},\,\ldots,\,
           e^{-j\pi(n_{\mathrm{t}}-1)\cos\phi_l}\bigr]^{\top},
  \label{eq:sv_tx} \\[4pt]
  \mathbf{a}_{\mathrm{r}}(\psi_l)
  &= \frac{1}{\sqrt{n_{\mathrm{r}}}}
     \bigl[1,\,e^{-j\pi\cos\psi_l},\,\ldots,\,
           e^{-j\pi(n_{\mathrm{r}}-1)\cos\psi_l}\bigr]^{\top}.
  \label{eq:sv_rx}
\end{align}
The complex path gains are modeled as i.i.d. complex Gaussian random variables, $\alpha_l \sim \mathcal{CN}(0,1/L)$, ensuring that the expected aggregate channel power is normalized to unity~\cite{tsdce}.
We consider a purely analog beamforming~(ABF) architecture, in line with the setting of~\cite{tsdce} that this work extends, and a pilot-based training phase carried out using Discrete Fourier Transform~(DFT)-based codebooks, in which the Tx sweeps over a codebook of $P$ beamforming vectors and the Rx over a codebook of $Q$ combining vectors, yielding $Q\times P$ direction pairs with quantized angles $\bar{\phi}_p$ and $\bar{\psi}_q$, respectively. Following~\cite{tsdce}, the codebook is designed so that the cosines of the quantized angles are uniformly spaced on $[-1,1]$:
\begin{align}
  \cos(\bar{\phi}_p)
  &= \mathcal{W}_{[-1,1]}\!\left(\frac{2p}{P}\right),
  \quad p = 0,\ldots,P-1,
  \label{eq:codebook_tx} \\[4pt]
  \cos(\bar{\psi}_q)
  &= \mathcal{W}_{[-1,1]}\!\left(-\frac{2q}{Q}\right),
  \quad q = 0,\ldots,Q-1,
  \label{eq:codebook_rx}
\end{align}
where $\mathcal{W}_{[a,b]}(x) \triangleq x-(b-a)\lceil\frac{x-b}{b-a}\rceil$ is the $[a,b]$ wrapping operator, with $\lceil\cdot\rceil$ denoting
the ceiling function. The beamforming and combining vectors are designed to match the array response~\cite{tsdce,Zhang2016}, i.e., $\mathbf{f}_p = \mathbf{a}_{\mathrm{t}}(\bar{\phi}_p)$ and $\mathbf{w}_q = \mathbf{a}_{\mathrm{r}}(\bar{\psi}_q)$, giving
$\mathbf{F} = [\mathbf{f}_0,\ldots,\mathbf{f}_{P-1}]
\in\mathbb{C}^{n_{\mathrm{t}}\times P}$ and
$\mathbf{W} = [\mathbf{w}_0,\ldots,\mathbf{w}_{Q-1}]
\in\mathbb{C}^{n_{\mathrm{r}}\times Q}$.
Assuming the transmission of the
pilot symbol $s=1$ through every Tx-Rx direction pair at each delay tap $d$, the observation matrix is:  %
\begin{equation}
    \mathbf{Y}[d] = \sqrt{\rho}\,\mathbf{W}^{\mathsf{H}} \mathbf{H}[d]\mathbf{F}
         + \mathbf{N}[d] = \sqrt{\rho}\sum_{l=1}^{L}\mathbf{G}^{(l)}[d]
         + \mathbf{N}[d],
  \label{eq:obs_matrix}
\end{equation}
where $\mathbf{Y}[d] \in \mathbb{C}^{Q \times P}$ and $\mathbf{G}^{(l)}[d]$ is the $l$-th path contribution:
\begin{equation}
  \mathbf{G}^{(l)}[d] =
  \sqrt{n_{\mathrm{t}} n_{\mathrm{r}}}\,\alpha_l
  \bigl(\mathbf{W}^{\mathsf{H}}\mathbf{a}_{\mathrm{r}}(\psi_l)\mathbf{a}_{\mathrm{t}}^{\mathsf{H}}(\phi_l)\mathbf{F}\bigr)
  p(d-\tau_l),
  \label{eq:path_contribution}
\end{equation}
$\rho\in\mathbb{R}^+$ is the pilot transmit power, and $\mathbf{N}[d]\in\mathbb{C}^{Q\times P}$ is additive white Gaussian noise with i.i.d.\ entries distributed as $\mathcal{CN}(0,\sigma_n^2)$, where $\sigma_n^2$ is the noise variance. Given the normalized channel, the system signal-to-noise ratio (SNR) is defined as $\rho/\sigma_n^2$.
The full wideband observation is a three-dimensional tensor $\mathcal{Y} \in \mathbb{C}^{Q\times P\times D}$, obtained by stacking the matrices $\mathbf{Y}[d]$ along the delay axis.
Similarly, $\mathcal{H} \in \mathbb{C}^{n_{\mathrm{r}} \times n_{\mathrm{t}} \times D}$ denotes the channel tensor obtained by stacking $\mathbf{H}[d]$ along the delay axis.
Due to the DFT-based codebook structure of~\eqref{eq:codebook_tx}--\eqref{eq:codebook_rx}, each slice $\mathbf{Y}[d]$ can be interpreted as the 2D-DFT of a sum of windowed complex sinusoids in the spatial domain, whose frequencies are directly related to the underlying AoDs and AoAs~\cite{tsdce}.

\section{The TSDCE-TD Algorithm}
\label{sec:algorithm}
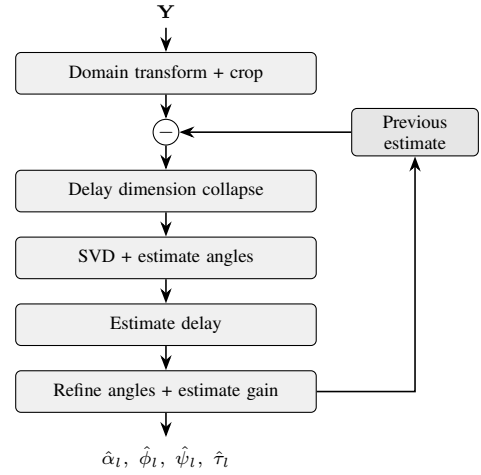
\begin{figure}[tb]
    \centering
    \centering
\begin{tikzpicture}[
    node distance=0.32cm,
    every node/.style={font=\scriptsize},
    block/.style={rectangle, rounded corners=2pt, draw, fill=gray!12,
                  text width=3.8cm, align=center,
                  minimum height=0.55cm, inner sep=2.5pt},
    prevb/.style={rectangle, rounded corners=2pt, draw, fill=gray!20,
                  text width=1.5cm, align=center,
                  minimum height=0.55cm, inner sep=2.5pt},
    sum/.style={circle, draw, inner sep=0pt, minimum size=0.34cm},  %
    arr/.style={-Stealth, semithick},
]
\node (Y) {$\mathbf{Y}$};
\node[block, below=0.32cm of Y] (dt)
    {Domain transform + crop};
\node[sum, below=0.32cm of dt] (sub) {$-$};  %
\node[block, below=0.32cm of sub] (dc)
    {Delay dimension collapse};
\node[block, below=0.32cm of dc] (ang)
    {SVD + estimate angles};
\node[block, below=0.32cm of ang] (del)
    {Estimate delay};
\node[block, below=0.32cm of del] (ref)
    {Refine angles + estimate gain};
\node[below=0.32cm of ref] (out)
    {$\hat\alpha_l,\;\hat\phi_l,\;\hat\psi_l,\;\hat\tau_l$};
\path (dt.east |- sub) ++(1.3cm,0)
    node[prevb] (prev) {Previous estimate};
\draw[arr] (Y)   -- (dt);
\draw[arr] (dt)  -- (sub);
\draw[arr] (sub) -- (dc);
\draw[arr] (dc)  -- (ang);
\draw[arr] (ang) -- (del);
\draw[arr] (del) -- (ref);
\draw[arr] (ref) -- (out);
\draw[arr] (prev.west) -- (sub.east);
\draw[arr] (ref.east)
    -- ++(0.5,0) coordinate (rc)
    -- (rc -| prev.south)
    -- (prev.south);
\end{tikzpicture}
    \caption{Block diagram of the TSDCE-TD algorithm.}
    \label{fig:tsdce-td-diagram}
\end{figure}

The proposed algorithm extends the narrowband TSDCE method of~\cite{tsdce} to the wideband setting by adding a delay estimation stage. %
Two variants are proposed: TSDCE-TD~(Dict), which searches a precomputed delay dictionary, and TSDCE-TD~(FFT), which achieves sub-tap accuracy via FFT zero-padding without any dictionary.
Both share the domain transform and successive interference cancellation~(SIC) skeleton described below and depicted in Fig.~\ref{fig:tsdce-td-diagram}.

\subsection{Domain Transform and Common Operations}
\label{ssec:common}

\subsubsection*{Spatial-domain transform}
Due to the DFT-codebook structure of $\mathbf{F}$ and $\mathbf{W}$~\eqref{eq:codebook_tx}--\eqref{eq:codebook_rx}, applying the inverse 2D-DFT (2D-IDFT), computed via the inverse FFT (IFFT), to each observation slice $\mathbf{Y}[d]$~\eqref{eq:obs_matrix} and retaining its top-left $n_{\mathrm{r}}\times n_{\mathrm{t}}$ block yields the \emph{spatial-domain matrix}:
\begin{equation}
  \mathbf{S}[d]
  = \bigl[\mathrm{IFFT}\bigl(\mathbf{Y}[d]\bigr)\bigr]_{1:n_{\mathrm{r}},\,1:n_{\mathrm{t}}}
  \in \mathbb{C}^{n_{\mathrm{r}} \times n_{\mathrm{t}}}.
  \label{eq:domain_transform}
\end{equation}

From~\eqref{eq:obs_matrix}--\eqref{eq:path_contribution}, each $\mathbf{S}[d]$ is approximately a sum of $L$ rank-1 terms $\sqrt{\rho}\,\alpha_l\,\mathbf{a}_{\mathrm{r}}(\psi_l)\mathbf{a}_{\mathrm{t}}^{\mathsf{H}}(\phi_l)\,p(d-\tau_l)$, whose spatial frequencies encode $(\phi_l,\psi_l)$ and are recovered by the weighted least-squares~(WLS) estimator of~\cite{tsdce}.

\subsubsection*{Delay template vector}
For any candidate delay $\tau$, define the $D$-dimensional \emph{delay template}:
\begin{equation}
  \mathbf{h}(\tau)
  = \bigl[p(0-\tau),\,p(1-\tau),\,\ldots,\,p(D-1-\tau)\bigr]^{\top},
  \label{eq:pulse_vec}
\end{equation}
where $p(\cdot)$ is the pulse-shaping filter of~\eqref{eq:channel_td}.
Since channel delay taps contain $p(d-\tau_l)$, projecting any observation-derived signal onto $\mathbf{h}(\tau)$ acts as a matched filter concentrating energy at $\tau\!\approx\!\tau_l$.

\subsubsection*{SIC residual}
Both variants run $K_{\mathrm ref}$ outer sweeps ($k=1,\ldots,K_{\mathrm ref}$), processing paths $l=1,\ldots,L$ in turn at each sweep. At step $(k,l)$, the contributions of all other paths are subtracted from $\mathbf{S}[d]$:
\begin{equation}
  \mathbf{S}'[d]
  = \mathbf{S}[d]
    - \sqrt{\rho}\sum_{j \neq l}\widehat{\mathbf{C}}_j[d],
  \label{eq:sic_residual}
\end{equation}
where
$\widehat{\mathbf{C}}_j[d] = \widehat\alpha_j\,\mathbf{a}_{\mathrm{r}}(\widehat\psi_j)\mathbf{a}_{\mathrm{t}}^{\mathsf{H}}(\widehat\phi_j)\, p(d-\widehat\tau_j)$
is the reconstructed spatial contribution of path~$j$.
All path contributions are initialized to $\widehat{\mathbf{C}}_j[d]=\mathbf{0}$ before the first sweep.

\subsubsection*{Spatial collapse}
Reshaping $\mathbf{S}'[d]$ into $\bar{\mathbf{S}}\in\mathbb{C}^{n_{\mathrm{r}}n_{\mathrm{t}}\times D}$ (columns indexed by delay tap) allows computing the \emph{spatial collapse} at a candidate delay $\widehat\tau_l$:
\begin{equation}
  \widehat{\mathbf{M}}
  = \operatorname{reshape}\bigl(
      \bar{\mathbf{S}}\,\mathbf{h}^*(\widehat\tau_l),\;n_{\mathrm{r}},\;n_{\mathrm{t}}
    \bigr)
  \in\mathbb{C}^{n_{\mathrm{r}}\times n_{\mathrm{t}}}.
  \label{eq:collapse}
\end{equation}
Matrix $\widehat{\mathbf{M}}$ concentrates the spatial signature of path $l$ into a single matrix, from which~\cite{tsdce} recovers $(\widehat\phi_l,\widehat\psi_l)$.
At $k\!=\!1$, $l\!<\!L$, a rank-1 singular value decomposition (SVD) truncation $\widehat{\mathbf{M}}\!\leftarrow\!\sigma_1\mathbf{u}_1\mathbf{v}_1^{\mathsf{H}}$ suppresses residual interference from yet-unestimated paths.

\subsubsection*{Delay profile and gain} After angle estimation, the residual is projected onto the estimated steering directions to obtain the scalar \emph{delay profile}:
\begin{equation}
  z[d]
  = \mathbf{a}_{\mathrm{r}}^{\mathsf{H}}(\widehat\psi_l)\,\mathbf{S}'[d]\,
    \mathbf{a}_{\mathrm{t}}(\widehat\phi_l),
  \quad d=0,\ldots,D-1.
  \label{eq:delay_profile}
\end{equation}
Writing $\mathbf{z}=[z[0],\ldots,z[D-1]]^{\top}$, the complex path gain follows by least squares:
\begin{equation}
  \widehat\alpha_l
  = \frac{\mathbf{h}^{\mathsf{H}}(\widehat\tau_l)\,\mathbf{z}}
         {\sqrt{\rho}\,\|\mathbf{h}(\widehat\tau_l)\|^2}.
  \label{eq:gain_ls}
\end{equation}
\subsection{Delay Estimation Variants}
\label{ssec:delay_variants}

\subsubsection{TSDCE-TD (Dict)}
This variant precomputes a dictionary of $G_d$ uniformly spaced delay templates~\eqref{eq:pulse_vec}:
\begin{equation}
  \mathbf{A}_\tau
  = \bigl[\mathbf{h}(\tau_1),\ldots,\mathbf{h}(\tau_{G_d})\bigr]
  \in\mathbb{R}^{D\times G_d}.
  \label{eq:delay_dict}
\end{equation}
An angle-free seed is obtained by power collapse across all $n_{\mathrm{r}}n_{\mathrm{t}}$ channels:
\begin{equation}
  \widehat\tau_l = \arg\max_{\tau_g}\,\bigl\|\bar{\mathbf{S}}\,\mathbf{h}^*(\tau_g)\bigr\|^2,
  \label{eq:power_collapse}
\end{equation}
and refined after angle estimation by $\widehat\tau_l\!\leftarrow\!\arg\max_{\tau_g}|\mathbf{z}^{\mathsf{H}}\mathbf{h}(\tau_g)|$.

\subsubsection{TSDCE-TD (FFT)}
This variant estimates $\widehat\tau_l$ via FFT-based interpolation with upsampling factor $N_{\mathrm up}$, bounding the delay error by $\Delta\tau\!\leq\!1/(2N_{\mathrm up})$ taps without any precomputed dictionary.
The angle-free seed uses the \emph{per-tap power}
\begin{equation}
  PW[d] = \bigl\|\mathbf{S}'[d]\bigr\|_F^2,
  \quad d=0,\ldots,D-1,
  \label{eq:power_scan}
\end{equation}
where $\|\cdot\|_F$ denotes the Frobenius norm.
The above metric concentrates near $d\!\approx\!\tau_l$, and its $N_{\mathrm{up}}$-point zero-padded IFFT yields the first-pass seed $\widehat\tau_l^{(1)}$.
Once the angles $(\widehat\phi_l,\widehat\psi_l)$ have been estimated, the delay is refined using the scalar delay profile $\mathbf{z}$~\eqref{eq:delay_profile}, which coherently combines all spatial channels via steering-vector projection and thus yields a sharper IFFT peak than the incoherent $PW[d]$.
Applying the same $N_{\mathrm{up}}$-point zero-padding to $\mathbf{z}$ and restricting the argmax to $\widehat\tau_l^{(1)}\!\pm\!1$ tap yields the refined estimate $\widehat\tau_l$.

Algorithm~\ref{alg:tsdce_td} shows the complete TSDCE-TD procedure for both proposed delay estimation variants.
\renewcommand{\algorithmicrequire}{\textbf{Input:}}
\renewcommand{\algorithmicensure}{\textbf{Output:}}
\begin{algorithm}[t]
\caption{TSDCE-TD (Dict / FFT)}
\label{alg:tsdce_td}
\begin{algorithmic}[1]
\Require $\mathcal{Y}$, $n_{\mathrm{r}}$, $n_{\mathrm{t}}$, $L$, $K_{\mathrm ref}$, $\rho$;\;
         \textbf{Dict}: $\mathbf{A}_\tau$~\eqref{eq:delay_dict};\;
         \textbf{FFT}: $N_{\mathrm up}$
\Ensure  $\{(\widehat\phi_l,\widehat\psi_l,\widehat\tau_l,\widehat\alpha_l)\}_{l=1}^L$
\State Domain transform: $\mathbf{S}[d]$~\eqref{eq:domain_transform}
       $\forall d$;\; init $\widehat{\mathbf{C}}_l[d]\!\leftarrow\!\mathbf{0}$
\For{$k = 1$ \textbf{to} $K_{\mathrm ref}$}
  \For{$l = 1$ \textbf{to} $L$}
    \State SIC residual $\mathbf{S}'[d]$~\eqref{eq:sic_residual},
           reshape to $\bar{\mathbf{S}}$
    \If{\textbf{(Dict)}}  \textit{angle-free seed~\eqref{eq:power_collapse}}
      \State $\widehat\tau_l \leftarrow
             \arg\max_{\tau_g} \|\bar{\mathbf{S}}\,\mathbf{h}^*(\tau_g)\|^2$
    \Else\ \textbf{(FFT)}
      \State $PW[d]\!=\!\|\mathbf{S}'[d]\|_F^2$~\eqref{eq:power_scan}
      \State $\widehat\tau_l^{(1)}\!\leftarrow\!\arg\max\,
             \mathrm{IFFT}_{N_{\mathrm up}}(PW)$
    \EndIf
    \State Spatial collapse $\widehat{\mathbf{M}}$~\eqref{eq:collapse}
    \If{$k\!=\!1$ \textbf{and} $l\!<\!L$}
      \State Rank-1 SVD:
             $\widehat{\mathbf{M}}\!\leftarrow\!\sigma_1\mathbf{u}_1\mathbf{v}_1^{\mathsf{H}}$
    \EndIf
    \State Angle estimate $(\widehat\phi_l,\widehat\psi_l)$:
           WLS on $\widehat{\mathbf{M}}$~\cite{tsdce}
    \State Delay profile $z[d]$~\eqref{eq:delay_profile}
    \If{\textbf{(Dict)}} \textit{matched-filter refinement}
      \State $\widehat\tau_l \leftarrow
             \arg\max_{\tau_g}\,|\mathbf{z}^{\mathsf{H}}\mathbf{h}(\tau_g)|$
    \Else\ \textbf{(FFT)}
      \State $\widehat\tau_l\!\leftarrow\!\arg\max_{|\tau-\widehat\tau_l^{(1)}|\leq 1}\,|\mathrm{IFFT}_{N_{\mathrm up}}(\mathbf{z})|$
    \EndIf
    \State Re-collapse $\widehat{\mathbf{M}}$~\eqref{eq:collapse};\;
           re-estimate $(\widehat\phi_l,\widehat\psi_l)$~\cite{tsdce}
    \State Gain $\widehat\alpha_l$~\eqref{eq:gain_ls};\;
           update $\widehat{\mathbf{C}}_l[d]$
  \EndFor
\EndFor
\end{algorithmic}
\end{algorithm}

\section{Performance Evaluation}
\label{sec:perf_evaluation}

Simulations are carried out using a wideband mmWave channel with $L = 3$ propagation paths at a carrier frequency of 60~GHz, corresponding to the minimum number of resolvable paths typically required for 2-D single-anchor localization with a ULA in the presence of an unknown clock offset~\cite{Shahmansoori2018}. The system operates over a bandwidth of $B = 2$~GHz, yielding a sampling period of $T_s = 0.5$~ns. The Tx and Rx are each equipped with $n_{\mathrm{t}} = n_{\mathrm{r}} = 16$ antennas, and the ABF codebooks contain $P = Q = 32$ codewords on each side. For channel generation, the pilot transmit power is set to $\rho = 1$ (i.e., the SNR is defined as $1/\sigma_n^2$), AoDs, AoAs, and path delays are drawn independently and uniformly from $[0,\pi]$, $[0,\pi]$, and $[0, N_\tau]$ samples, respectively. The delay axis consists of $N_\tau = 64$ base delay taps, extended by $n_{\mathrm{e}} = 8$ early guard samples and $n_{\mathrm{l}} = 8$ late guard samples, yielding a total of $D = N_\tau + n_{\mathrm{e}} + n_{\mathrm{l}} = 80$ delay taps per observation slice.
The pulse shaping filter is a pure sinc function~\cite{momp}.

The proposed TSDCE-TD methods are benchmarked against OMP and Multidimensional OMP~(MOMP)~\cite{momp}. MOMP is an OMP-based method that reduces complexity by replacing the joint dictionary with a set of independent per-dimension dictionaries.
To configure these algorithms, an oversampling factor $K_{\mathrm{res}} \geq 1$ scales both the angle and delay dictionaries simultaneously: it defines a delay grid of $G_d = N_\tau K_{\mathrm{res}}$ candidates and, for the OMP-based baselines, angular grids of $G_p = P K_{\mathrm{res}}$ codewords per Tx/Rx dimension. Both baseline implementations and their specific parameters follow the original work~\cite{momp}.

OMP and MOMP are evaluated with $K_{\mathrm{res}} \in \{2, 4, 8\}$;
larger values exceed the available memory capacity of our system, equipped with an Intel i9-9940X CPU, 128~GB of RAM, and an NVIDIA RTX Titan GPU, due to OMP's joint dictionary size.
TSDCE-TD~(Dict) uses the same value of $K_{\mathrm{res}} = 8$ for the delay axis only; TSDCE-TD~(FFT) replaces the dictionary with a zero-padded FFT ($N_{\mathrm{up}} = 8$). For TSDCE-TD variants, $K_{\mathrm{ref}} = 2$ outer refinement iterations are used.

\subsection{Channel Estimation Performance}
Channel estimation performance is assessed via the normalized mean squared error~(NMSE) of the reconstructed wideband channel tensor, given by:
\begin{equation}
    \mathrm{NMSE}\,[\mathrm{dB}]
    = 10\log_{10}\!\left(
        \mathbb{E}\!\left\{
            \frac{\|\widehat{\mathcal{H}} - \mathcal{H}\|_F^2}
                 {\|\mathcal{H}\|_F^2}
        \right\}
      \right),
    \label{eq:nmse}
\end{equation}
\noindent where $\mathcal{H}$ and $\widehat{\mathcal{H}}$ denote
the true and estimated channel tensors, respectively.
In addition, we evaluate the root mean square error~(RMSE) of the AoD and AoA estimates computed on the detected paths. The angle RMSE is defined as:
\begin{equation}
    \mathrm{RMSE}_\vartheta
    = \sqrt{\frac{1}{|\mathcal{N}|}
        \sum_{i \in \mathcal{N}}
        \left(\vartheta_i - \widehat{\vartheta}_i\right)^2},
    \label{eq:rmse_angle}
\end{equation}
where $\mathcal{N}$ is the set of detected paths and $\vartheta_i$, $\widehat{\vartheta}_i$ denote the true and estimated angles (either AoA or AoD) of the $i$-th path, respectively.
A path is considered detected if its estimated angles fall within $1^\circ$ of the true value; the probability of detection ($P_{\mathrm d}$) reflects the fraction of trials in which paths satisfy this criterion. Finally, delay accuracy is assessed through the empirical cumulative distribution function (CDF) of the delay estimation error in nanoseconds (ns) at a fixed SNR. For this metric, the estimated delays are matched to the ground-truth delays using one-dimensional nearest-neighbor matching, independently of the angle-detection criterion.
Simulations are averaged over 1000 independent channel realizations for SNR values ranging from $-10\text{ dB}$ to $25\text{ dB}$, following the observation model in~\eqref{eq:obs_matrix}.

\begin{figure}[tb]
    \centering
    \centering
\begin{tikzpicture}
\begin{axis}[
  ieee single col,
  width=0.95\linewidth,
  height=7cm,
  xmin=-10, xmax=25,
  ymin=-38.5, ymax=0,
  xtick={-10,-5,0,5,10,15,20,25},
  ytick={-35,-30,-25,-20,-15,-10,-5,0},
  axis lines=box,
  tick pos=left,
  xlabel={SNR [dB]},
  ylabel={NMSE [dB]},
   legend style={at={(0.02,0.02)}, anchor=south west, font=\fontsize{5}{6}\selectfont},
]

\addplot[
  color=colorTSDCEfft, solid, line width=0.9pt,
  mark=diamond*, mark size=2pt,
  mark options={solid,fill=colorTSDCEfft,draw=colorTSDCEfft},
] coordinates {
  (-10,-2.9003)
  (-5,-9.4488)
  (0,-16.4777)
  (5,-21.1194)
  (10,-23.4149)
  (15,-24.4733)
  (20,-24.7077)
  (25,-24.9903)
};
\addlegendentry{TSDCE-TD (FFT)}

\addplot[
  color=colorTSDCEdict, solid, line width=0.9pt,
  mark=*, mark size=2pt,
  mark options={solid,fill=colorTSDCEdict,draw=colorTSDCEdict},
] coordinates {
  (-10,-3.0560)
  (-5,-9.9709)
  (0,-16.9018)
  (5,-21.1420)
  (10,-23.6363)
  (15,-24.3807)
  (20,-24.8140)
  (25,-24.9408)
};
\addlegendentry{TSDCE-TD (Dict)}

\addplot[
  color=colorOMP, solid, line width=0.9pt,
  mark=square*, mark size=2pt,
  mark options={solid,fill=colorOMP,draw=colorOMP},
] coordinates {
  (-10,-6.5810)
  (-5,-12.8201)
  (0,-17.5690)
  (5,-20.1795)
  (10,-21.6115)
  (15,-21.9463)
  (20,-22.1270)
  (25,-22.2043)
};
\addlegendentry{OMP $K_{\rm res}=8$}

\addplot[
  color=colorOMP, dashed, line width=0.9pt,
  mark=square*, mark size=2pt,
  mark options={solid,fill=colorOMP,draw=colorOMP},
] coordinates {
  (-10,-5.9551)
  (-5,-11.3027)
  (0,-14.3738)
  (5,-15.6829)
  (10,-16.0469)
  (15,-16.2067)
  (20,-16.2812)
  (25,-16.2894)
};
\addlegendentry{OMP $K_{\rm res}=4$}

\addplot[
  color=colorOMP, dotted, line width=0.9pt,
  mark=square*, mark size=2pt,
  mark options={solid,fill=colorOMP,draw=colorOMP},
] coordinates {
  (-10,-5.0237)
  (-5,-8.7982)
  (0,-10.2750)
  (5,-10.6662)
  (10,-10.8716)
  (15,-11.0267)
  (20,-11.0164)
  (25,-10.9854)
};
\addlegendentry{OMP $K_{\rm res}=2$}

\addplot[
  color=colorMOMP, solid, line width=0.9pt,
  mark=triangle*, mark size=2pt,
  mark options={solid,fill=colorMOMP,draw=colorMOMP},
] coordinates {
  (-10,-2.1712)
  (-5,-8.1528)
  (0,-14.9991)
  (5,-19.1461)
  (10,-21.1819)
  (15,-21.8259)
  (20,-21.9681)
  (25,-22.1967)
};
\addlegendentry{MOMP $K_{\rm res}=8$}

\addplot[
  color=colorMOMP, dashed, line width=0.9pt,
  mark=triangle*, mark size=2pt,
  mark options={solid,fill=colorMOMP,draw=colorMOMP},
] coordinates {
  (-10,-2.0290)
  (-5,-7.3056)
  (0,-12.4218)
  (5,-14.7975)
  (10,-15.8115)
  (15,-16.0925)
  (20,-16.0873)
  (25,-16.0529)
};
\addlegendentry{MOMP $K_{\rm res}=4$}

\addplot[
  color=colorMOMP, dotted, line width=0.9pt,
  mark=triangle*, mark size=2pt,
  mark options={solid,fill=colorMOMP,draw=colorMOMP},
] coordinates {
  (-10,-1.3806)
  (-5,-5.0148)
  (0,-8.5312)
  (5,-9.8841)
  (10,-10.2639)
  (15,-10.2874)
  (20,-10.3663)
  (25,-10.4759)
};
\addlegendentry{MOMP $K_{\rm res}=2$}

\addplot[black, densely dotted, line width=0.7pt, forget plot] coordinates {
  (25,-21.4005)
  (24.9443,-21.4011)
  (24.8888,-21.4030)
  (24.8334,-21.4062)
  (24.7782,-21.4106)
  (24.7235,-21.4163)
  (24.6691,-21.4232)
  (24.6153,-21.4313)
  (24.5621,-21.4406)
  (24.5096,-21.4512)
  (24.4578,-21.4629)
  (24.4069,-21.4758)
  (24.357,-21.4899)
  (24.3081,-21.5050)
  (24.2602,-21.5213)
  (24.2135,-21.5387)
  (24.1681,-21.5570)
  (24.124,-21.5765)
  (24.0813,-21.5969)
  (24.04,-21.6182)
  (24.0003,-21.6405)
  (23.9621,-21.6636)
  (23.9256,-21.6876)
  (23.8907,-21.7124)
  (23.8576,-21.7380)
  (23.8264,-21.7643)
  (23.7969,-21.7914)
  (23.7694,-21.8190)
  (23.7439,-21.8473)
  (23.7203,-21.8761)
  (23.6987,-21.9054)
  (23.6792,-21.9352)
  (23.6618,-21.9654)
  (23.6465,-21.9960)
  (23.6334,-22.0269)
  (23.6224,-22.0581)
  (23.6135,-22.0895)
  (23.6069,-22.1211)
  (23.6025,-22.1528)
  (23.6003,-22.1846)
  (23.6003,-22.2164)
  (23.6025,-22.2482)
  (23.6069,-22.2799)
  (23.6135,-22.3115)
  (23.6224,-22.3429)
  (23.6334,-22.3741)
  (23.6465,-22.4050)
  (23.6618,-22.4356)
  (23.6792,-22.4658)
  (23.6987,-22.4956)
  (23.7203,-22.5249)
  (23.7439,-22.5537)
  (23.7694,-22.5820)
  (23.7969,-22.6097)
  (23.8264,-22.6367)
  (23.8576,-22.6630)
  (23.8907,-22.6886)
  (23.9256,-22.7134)
  (23.9621,-22.7374)
  (24.0003,-22.7605)
  (24.04,-22.7828)
  (24.0813,-22.8042)
  (24.124,-22.8245)
  (24.1681,-22.8440)
  (24.2135,-22.8623)
  (24.2602,-22.8797)
  (24.3081,-22.8960)
  (24.357,-22.9111)
  (24.4069,-22.9252)
  (24.4578,-22.9381)
  (24.5096,-22.9498)
  (24.5621,-22.9604)
  (24.6153,-22.9697)
  (24.6691,-22.9778)
  (24.7235,-22.9847)
  (24.7782,-22.9904)
  (24.8334,-22.9948)
  (24.8888,-22.9980)
  (24.9443,-22.9999)
  (25,-23.0005)
};
\node[anchor=south east, font=\fontsize{7}{8}\selectfont, text=black]
  at (axis cs:25,-22) {$K_{\rm res}=8$};

\addplot[black, densely dotted, line width=0.7pt, forget plot] coordinates {
  (25,-15.3711)
  (24.9443,-15.3718)
  (24.8888,-15.3737)
  (24.8334,-15.3768)
  (24.7782,-15.3812)
  (24.7235,-15.3869)
  (24.6691,-15.3938)
  (24.6153,-15.4019)
  (24.5621,-15.4113)
  (24.5096,-15.4218)
  (24.4578,-15.4336)
  (24.4069,-15.4465)
  (24.357,-15.4605)
  (24.3081,-15.4757)
  (24.2602,-15.4919)
  (24.2135,-15.5093)
  (24.1681,-15.5277)
  (24.124,-15.5471)
  (24.0813,-15.5675)
  (24.04,-15.5888)
  (24.0003,-15.6111)
  (23.9621,-15.6343)
  (23.9256,-15.6583)
  (23.8907,-15.6831)
  (23.8576,-15.7087)
  (23.8264,-15.7350)
  (23.7969,-15.7620)
  (23.7694,-15.7896)
  (23.7439,-15.8179)
  (23.7203,-15.8467)
  (23.6987,-15.8760)
  (23.6792,-15.9058)
  (23.6618,-15.9361)
  (23.6465,-15.9666)
  (23.6334,-15.9976)
  (23.6224,-16.0287)
  (23.6135,-16.0601)
  (23.6069,-16.0917)
  (23.6025,-16.1234)
  (23.6003,-16.1552)
  (23.6003,-16.1870)
  (23.6025,-16.2188)
  (23.6069,-16.2505)
  (23.6135,-16.2821)
  (23.6224,-16.3135)
  (23.6334,-16.3447)
  (23.6465,-16.3756)
  (23.6618,-16.4062)
  (23.6792,-16.4364)
  (23.6987,-16.4662)
  (23.7203,-16.4956)
  (23.7439,-16.5244)
  (23.7694,-16.5526)
  (23.7969,-16.5803)
  (23.8264,-16.6073)
  (23.8576,-16.6336)
  (23.8907,-16.6592)
  (23.9256,-16.6840)
  (23.9621,-16.7080)
  (24.0003,-16.7312)
  (24.04,-16.7534)
  (24.0813,-16.7748)
  (24.124,-16.7952)
  (24.1681,-16.8146)
  (24.2135,-16.8330)
  (24.2602,-16.8503)
  (24.3081,-16.8666)
  (24.357,-16.8818)
  (24.4069,-16.8958)
  (24.4578,-16.9087)
  (24.5096,-16.9204)
  (24.5621,-16.9310)
  (24.6153,-16.9403)
  (24.6691,-16.9485)
  (24.7235,-16.9554)
  (24.7782,-16.9610)
  (24.8334,-16.9655)
  (24.8888,-16.9686)
  (24.9443,-16.9705)
  (25,-16.9711)
};
\node[anchor=south east, font=\fontsize{7}{8}\selectfont, text=black]
  at (axis cs:25,-16) {$K_{\rm res}=4$};

\addplot[black, densely dotted, line width=0.7pt, forget plot] coordinates {
  (25,-9.8759)
  (24.9443,-9.8765)
  (24.8888,-9.8786)
  (24.8334,-9.8819)
  (24.7782,-9.8867)
  (24.7235,-9.8927)
  (24.6691,-9.9001)
  (24.6153,-9.9088)
  (24.5621,-9.9188)
  (24.5096,-9.9300)
  (24.4578,-9.9426)
  (24.4069,-9.9563)
  (24.357,-9.9714)
  (24.3081,-9.9876)
  (24.2602,-10.0050)
  (24.2135,-10.0235)
  (24.1681,-10.0431)
  (24.124,-10.0639)
  (24.0813,-10.0857)
  (24.04,-10.1085)
  (24.0003,-10.1323)
  (23.9621,-10.1570)
  (23.9256,-10.1827)
  (23.8907,-10.2092)
  (23.8576,-10.2365)
  (23.8264,-10.2646)
  (23.7969,-10.2935)
  (23.7694,-10.3230)
  (23.7439,-10.3532)
  (23.7203,-10.3840)
  (23.6987,-10.4154)
  (23.6792,-10.4472)
  (23.6618,-10.4795)
  (23.6465,-10.5122)
  (23.6334,-10.5452)
  (23.6224,-10.5785)
  (23.6135,-10.6121)
  (23.6069,-10.6458)
  (23.6025,-10.6797)
  (23.6003,-10.7137)
  (23.6003,-10.7476)
  (23.6025,-10.7816)
  (23.6069,-10.8155)
  (23.6135,-10.8492)
  (23.6224,-10.8828)
  (23.6334,-10.9161)
  (23.6465,-10.9492)
  (23.6618,-10.9818)
  (23.6792,-11.0141)
  (23.6987,-11.0460)
  (23.7203,-11.0773)
  (23.7439,-11.1081)
  (23.7694,-11.1383)
  (23.7969,-11.1678)
  (23.8264,-11.1967)
  (23.8576,-11.2248)
  (23.8907,-11.2521)
  (23.9256,-11.2787)
  (23.9621,-11.3043)
  (24.0003,-11.3290)
  (24.04,-11.3528)
  (24.0813,-11.3757)
  (24.124,-11.3974)
  (24.1681,-11.4182)
  (24.2135,-11.4378)
  (24.2602,-11.4564)
  (24.3081,-11.4737)
  (24.357,-11.4900)
  (24.4069,-11.5050)
  (24.4578,-11.5187)
  (24.5096,-11.5313)
  (24.5621,-11.5426)
  (24.6153,-11.5525)
  (24.6691,-11.5612)
  (24.7235,-11.5686)
  (24.7782,-11.5747)
  (24.8334,-11.5794)
  (24.8888,-11.5827)
  (24.9443,-11.5848)
  (25,-11.5854)
};
\node[anchor=south east, font=\fontsize{7}{8}\selectfont, text=black]
  at (axis cs:25,-10.5) {$K_{\rm res}=2$};

\end{axis}
\end{tikzpicture} %
    \caption{Channel NMSE vs.\ SNR; $L=3$, $n_{\mathrm{t}}=n_{\mathrm{r}}=16$.}
    \label{fig:nmse}
\end{figure}
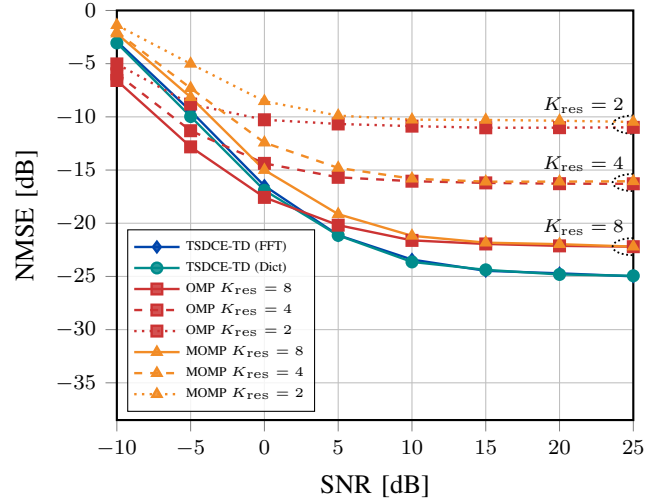

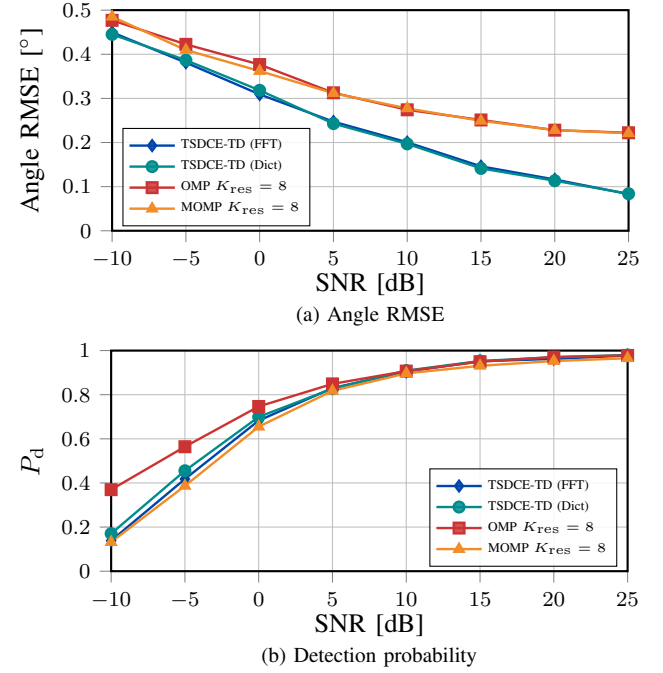
\begin{figure}[tb]
  \centering
  \begin{minipage}{\linewidth}
    \centering
\begin{tikzpicture}[baseline=(current bounding box.north)]
\begin{axis}[
  ieee single col,
  width=0.95\linewidth,
  height=4.5cm,
  xmin=-10, xmax=25,
  ymin=0, ymax=0.5,
  xtick={-10,-5,0, 5,10, 15,20, 25},
  ytick={0,0.1,0.2,0.3,0.4,0.5},
  xlabel={SNR [dB]},
  ylabel={Angle RMSE [$^\circ$]},
  axis lines=box,
  tick pos=left,
  title={(a) Angle RMSE},
  xlabel style={
  at={(axis description cs:0.5,-0.15)},
  anchor=north
},
title style={
  at={(axis description cs:0.5,-0.375)},
  anchor=north
},
legend style={
  at={(0.02,0.02)},
  anchor=south west,
  font=\fontsize{5}{6}\selectfont
},
]

\addplot[
  color=colorTSDCEfft, solid, line width=0.9pt,
  mark=diamond*, mark size=2pt,
  mark options={solid,fill=colorTSDCEfft,draw=colorTSDCEfft},
] coordinates {
  (-10,0.4497)
  (-5,0.3811)
  (0,0.3091)
  (5,0.2473)
  (10,0.2005)
  (15,0.1460)
  (20,0.1162)
  (25,0.0831)
};
\addlegendentry{TSDCE-TD (FFT)}

\addplot[
  color=colorTSDCEdict, solid, line width=0.9pt,
  mark=*, mark size=2pt,
  mark options={solid,fill=colorTSDCEdict,draw=colorTSDCEdict},
] coordinates {
  (-10,0.4451)
  (-5,0.3866)
  (0,0.3183)
  (5,0.2429)
  (10,0.1969)
  (15,0.1413)
  (20,0.1134)
  (25,0.0839)
};
\addlegendentry{TSDCE-TD (Dict)}

\addplot[
  color=colorOMP, solid, line width=0.9pt,
  mark=square*, mark size=2pt,
  mark options={solid,fill=colorOMP,draw=colorOMP},
] coordinates {
  (-10,0.4769)
  (-5,0.4225)
  (0,0.3767)
  (5,0.3127)
  (10,0.2741)
  (15,0.2511)
  (20,0.2280)
  (25,0.2221)
};
\addlegendentry{OMP $K_{\mathrm{res}}=8$}

\addplot[
  color=colorMOMP, solid, line width=0.9pt,
  mark=triangle*, mark size=2pt,
  mark options={solid,fill=colorMOMP,draw=colorMOMP},
] coordinates {
  (-10,0.4850)
  (-5,0.4097)
  (0,0.3620)
  (5,0.3117)
  (10,0.2774)
  (15,0.2492)
  (20,0.2282)
  (25,0.2210)
};
\addlegendentry{MOMP $K_{\mathrm{res}}=8$}

\end{axis}
\end{tikzpicture}
  \end{minipage}

  \begin{minipage}{\linewidth}
    \centering
\begin{tikzpicture}[baseline=(current bounding box.north)]
\begin{axis}[
  ieee single col,
  width=0.95\linewidth,
  height=4.5cm,
  xmin=-10, xmax=25,
  ymin=0, ymax=1,
  xtick={-10,-5,0, 5,10, 15,20, 25},
  ytick={0,0.2,0.4,0.6,0.8,1.0},
  xlabel={SNR [dB]},
  ylabel={$P_{\mathrm d}$},
  axis lines=box,
  tick pos=left,
  title={(b) Detection probability},
  xlabel style={
  at={(axis description cs:0.5,-0.15)},
  anchor=north
},
title style={
  at={(axis description cs:0.5,-0.375)},
  anchor=north
},
  legend style={at={(0.98,0.02)}, anchor=south east, font=\fontsize{5}{6}\selectfont},
]

\addplot[
  color=colorTSDCEfft, solid, line width=0.9pt,
  mark=diamond*, mark size=2pt,
  mark options={solid,fill=colorTSDCEfft,draw=colorTSDCEfft},
] coordinates {
  (-10,0.1383)
  (-5,0.4177)
  (0,0.6827)
  (5,0.8300)
  (10,0.9047)
  (15,0.9517)
  (20,0.9623)
  (25,0.9777)
};
\addlegendentry{TSDCE-TD (FFT)}

\addplot[
  color=colorTSDCEdict, solid, line width=0.9pt,
  mark=*, mark size=2pt,
  mark options={solid,fill=colorTSDCEdict,draw=colorTSDCEdict},
] coordinates {
  (-10,0.1703)
  (-5,0.4553)
  (0,0.6993)
  (5,0.8273)
  (10,0.9093)
  (15,0.9527)
  (20,0.9700)
  (25,0.9800)
};
\addlegendentry{TSDCE-TD (Dict)}

\addplot[
  color=colorOMP, solid, line width=0.9pt,
  mark=square*, mark size=2pt,
  mark options={solid,fill=colorOMP,draw=colorOMP},
] coordinates {
  (-10,0.3697)
  (-5,0.5643)
  (0,0.7467)
  (5,0.8490)
  (10,0.9070)
  (15,0.9497)
  (20,0.9707)
  (25,0.9780)
};
\addlegendentry{OMP $K_{\rm res}=8$}

\addplot[
  color=colorMOMP, solid, line width=0.9pt,
  mark=triangle*, mark size=2pt,
  mark options={solid,fill=colorMOMP,draw=colorMOMP},
] coordinates {
  (-10,0.1320)
  (-5,0.3867)
  (0,0.6550)
  (5,0.8177)
  (10,0.8967)
  (15,0.9310)
  (20,0.9527)
  (25,0.9653)
};
\addlegendentry{MOMP $K_{\rm res}=8$}

\end{axis}
\end{tikzpicture}
  \end{minipage}
  \caption{Angle RMSE and probability of detection vs.\ SNR, $L=3$, $n_{\mathrm{t}}=n_{\mathrm{r}}=16$.}
  \label{fig:angle_pd}
\end{figure}

Fig.~\ref{fig:nmse} shows the channel NMSE versus SNR. %
Both variants of TSDCE-TD outperform OMP and MOMP at medium-to-high SNR. At low SNR, OMP benefits from jointly correlating all dimensions simultaneously, making it more robust under high noise.
Fig.~\ref{fig:angle_pd} shows the angle RMSE and $P_{\mathrm d}$. The TSDCE-TD methods consistently achieve lower angle RMSE, but they only attain a comparable $P_{\mathrm{d}}$ to OMP at medium-to-high SNRs.
Finally, Fig.~\ref{fig:delay_cdf} shows the empirical CDF of the per-path delay error
at $\mathrm{SNR}=10\text{ dB}$.
Both TSDCE-TD variants achieve delay accuracy comparable to OMP and MOMP
at $K_{\mathrm{res}}=8$.

Overall, the proposed TSDCE-TD methods achieve very similar channel estimation performance to OMP and MOMP with $K_{\mathrm{res}}=8$, while outperforming both in channel reconstruction and angle estimation at medium to high SNR.
While performance can be further improved by increasing $K_{\mathrm{res}}$ or $N_{\mathrm{up}}$ at a fraction of the baselines' complexity, the proposed methods remain limited at low SNR by separate parameter estimation and SIC error propagation.
\begin{figure}[tb]
  \input{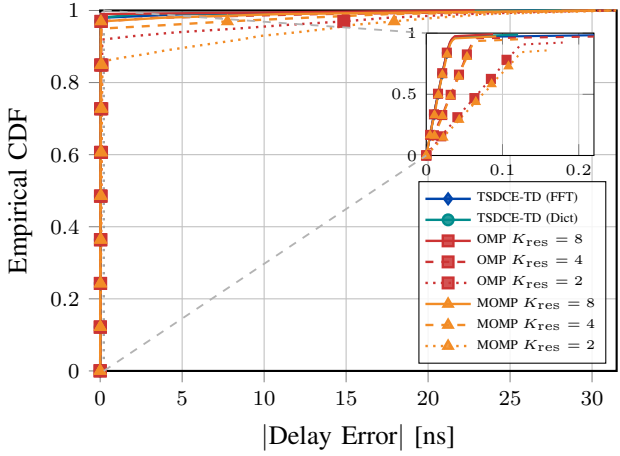}
  \caption{Empirical CDF of the delay error at SNR = 10~dB; $L=3$, $n_{\mathrm{t}}=n_{\mathrm{r}}=16$. The top-right inset focuses on the $0$--$0.22$~ns delay region.}
  \label{fig:delay_cdf}
\end{figure}

\subsection{Localization Performance}
\label{ssec:localization}
\begin{figure}[tb]
    \centering
    \includegraphics[width=0.65\columnwidth]{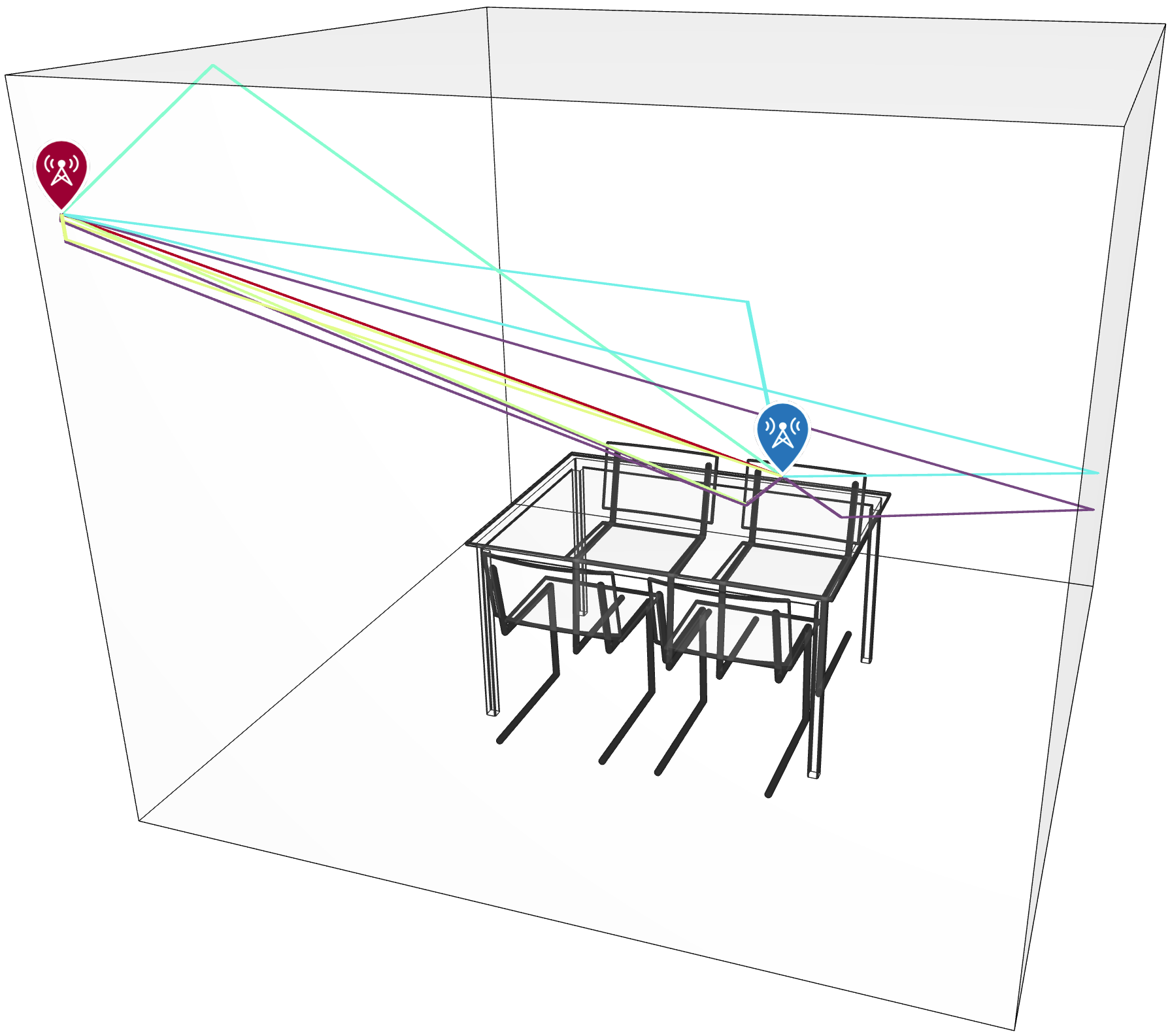}
    \caption{Ray-tracing scenario used for localization evaluation.
             Rx in red; one representative Tx position in blue. Ten propagation rays are considered.}
    \label{fig:indoor_scenario}
\end{figure}

Beyond communication-oriented channel reconstruction metrics, the estimated angular and delay parameters are also directly relevant for sensing and localization tasks in ISAC systems. In particular, accurate AoA/AoD and delay estimation enables geometric inference of user position from multipath propagation.
The localization capability of TSDCE-TD is assessed in the indoor ray-tracing scenario illustrated in Fig.~\ref{fig:indoor_scenario}, which represents a furnished room with concrete walls and wooden furniture, provided by the MATLAB \texttt{conferenceroom.stl} model.

The receiver is a static anchor mounted at the ceiling corner at location $\mathbf{p}_\mathrm{r} = [x_{\mathrm{r}},y_{\mathrm{r}},z_{\mathrm{r}}]^{\top} = [\mathbf{p}_{\mathrm{r}}^{(xy)},z_{\mathrm{r}}]^{\top}$, and equipped with a 16-element ULA. The same waveform parameters used in the synthetic-data experiments are applied here, and the user equipment (Tx) is identically configured.
Tx positions are drawn from a uniform $32{\times}32$ grid over the area $[-1.4,\,1.4]\times [-1.4,\,1.4]$\,m$^2$ at a fixed height of $0.85$~m. Each 3-D Tx location is identified by $\mathbf{p}_{\mathrm{t}} = [x_{\mathrm{t}},y_{\mathrm{t}},z_{\mathrm{t}}]^{\top} = [\mathbf{p}_{\mathrm{t}}^{(xy)},z_{\mathrm{t}}]^{\top} $.
The multipath channel is generated via ray tracing with up to two specular reflections, retaining the $L = 10$ strongest paths.
Channel gains are normalized to unit total power.
Under the assumption of perfect synchronization, the propagation delay of the
$\ell$-th estimated path is
\begin{equation}
    \widehat{\tau}_\ell = \tau_\mathrm{ref} + \widehat{d}_\ell \, T_s ,
    \label{eq:abs_delay}
\end{equation}
where $\tau_\mathrm{ref}$ is the reference (minimum) delay of the channel, $\widehat{d}_\ell$ is the estimated delay in samples, and $T_s$ is the sampling period.
Position estimation follows a single-anchor bearing-and-range approach applied to the dominant path, which at 60~GHz reasonably corresponds to the line-of-sight component~\cite{Shahmansoori2018}; it is identified as the path with the largest estimated gain magnitude:
\begin{equation}
    \ell^* = \arg\max_{\ell}\,|\widehat{\alpha}_\ell| .
    \label{eq:path_select}
\end{equation}
Given the selected AoA $\widehat{\psi} \triangleq \widehat{\psi}_{\ell^*}$ and
delay $\widehat{\tau} \triangleq \widehat{\tau}_{\ell^*}$, the corresponding 3-D propagation distance (slant range) is $\widehat{r}_{3\mathrm{D}} = c\widehat{\tau}$ with $c$ the speed of light, and the horizontal range is recovered using the known vertical separation $\Delta z = z_\mathrm{r} - z_\mathrm{t}$:
\begin{equation}
    \widehat{r}_{2\mathrm{D}} = \sqrt{\widehat{r}_{3\mathrm{D}}^2 - \Delta z^2} .
    \label{eq:r2d}
\end{equation}
The ULA response encodes the 3-D spatial frequency $\cos\widehat{\psi} = (x_\mathrm{t} - x_\mathrm{r})/\widehat{r}_{3\mathrm{D}}$, leading to the horizontal displacement of the user along the array axis
\begin{equation}
    \widehat{p}_\parallel = \widehat{r}_{3\mathrm{D}} \cos\widehat{\psi} ,
    \label{eq:proj}
\end{equation}
and the perpendicular component is
$\widehat{p}_\perp = \sqrt{\widehat{r}_{2\mathrm{D}}^2 - \widehat{p}_\parallel^2}$,
with the sign fixed by the known Rx orientation (the array faces into the room interior).
The Tx estimated position is
\begin{equation}
    \widehat{\mathbf{p}}_{\mathrm{t}}^{(xy)}
        = \mathbf{p}_\mathrm{r}^{(xy)}
          + \widehat{p}_\parallel\,\mathbf{u}
          + \widehat{p}_\perp\,\mathbf{u}_\perp ,
    \label{eq:pos_est}
\end{equation}
where $\mathbf{u}$ is the unit vector along the ULA axis and $\mathbf{u}_\perp$ its in-plane
perpendicular pointing into the room.

\begin{figure}[t]
    \centering
    \input{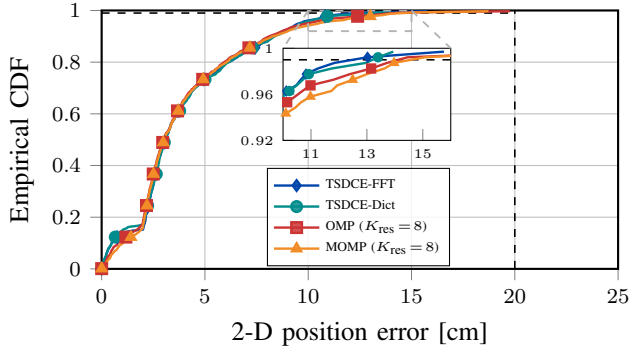} %
    \caption{Empirical CDF of the 2-D position error; SNR\,=\,10\,dB. The dashed lines mark the ETSI 99\textsuperscript{th}-percentile accuracy requirement of $20$\,cm~\cite{ETSI22261}.
    }
    \label{fig:pos_cdf}
\end{figure}

Defining the 2-D error of Tx position as $e_{\mathrm{t}} = \bigl\|\widehat{\mathbf{p}}_{\mathrm{t}}^{(xy)} - \mathbf{p}_{\mathrm{t}}^{(xy)}\bigr\|_2$, in Fig.~\ref{fig:pos_cdf} we report its CDF aggregated over 1000 realizations at a fixed SNR of 10~dB in the 1024 sample locations.
The CDF format is adopted according to the reporting convention of the ETSI indoor positioning specifications~\cite{ETSI22261}, whose Table~7.3.2.2-1 sets a horizontal positioning requirement of $20$\,cm at the 99\textsuperscript{th} percentile; this target is marked in Fig.~\ref{fig:pos_cdf} for direct comparison.

The results confirm that the proposed estimators satisfy the ETSI requirements for target positioning for 5G commercial indoor use cases~\cite{ETSI22261}.
The positioning accuracy obtained with TSDCE-TD is consistent with values reported in the
literature~\cite{SUN2024104372, Shahmansoori2018}, confirming that the channel-parameter estimates carry sufficient geometric information for practical localization.

\subsection{Computational Complexity}
\label{ssec:complexity}

We analyze the per-estimate complexity of the three algorithms, using $G_d$, $G_p$ for the delay and angle grid sizes, and $G_{p,\mathrm{lr}}$, $G_{d,\mathrm{lr}}$ for the corresponding low-resolution grids of MOMP.

\subsubsection{TSDCE-TD}
Angle and delay are estimated sequentially and independently, so the dominant cost scales with $n_{\mathrm{r}} n_{\mathrm{t}}$ and $G_d$ separately.

\subsubsection{OMP}
OMP must evaluate all $G_p^2 G_d$ joint atoms by forming the full Kronecker-product dictionary, giving a cost of $\mathcal{O}(G_p^2\,G_d\cdot QP\cdot D)$ that grows as $K_{\mathrm{res}}^3$.

\subsubsection{MOMP}
MOMP introduces low-resolution grids $G_{p,\mathrm{lr}}\!=\!P K_{\mathrm{res,lr}}$ and $G_{d,\mathrm{lr}}\!=\!N_\tau K_{\mathrm{res,lr}}$ (with $K_{\mathrm{res,lr}}$ following~\cite{momp}) for an initial greedy stage, followed by per-dimension refinement, substantially reducing the joint search cost relative to OMP.

Table~\ref{tab:complexity} lists the dominant-term formula and the evaluated cost for the simulation setup above ($n_{\mathrm{t}}\!=\!n_{\mathrm{r}}\!=\!16$, $P\!=\!Q\!=\!32$, $N_\tau\!=\!64$, $D\!=\!80$) at $K_{\mathrm{res}}\!=\!8$ ($G_p\!=\!256$, $G_d\!=\!512$, $G_{p,\mathrm{lr}}\!=\!128$, $G_{d,\mathrm{lr}}\!=\!256$, $N_{\mathrm{ref}}\!=\!3$, $K_{\mathrm{ref}} = 2$). MOMP and OMP are approximately four and six orders of magnitude more complex than TSDCE-TD, respectively.

\begin{table}[tb]
    \centering
    \caption{Dominant complexity per path estimate and evaluated cost for the setup
             in Section~\ref{sec:perf_evaluation} with $K_{\mathrm{res}}\!=\!8$.}
    \label{tab:complexity}
    \resizebox{\columnwidth}{!}{%
    \begin{tabular}{lll}
        \toprule
        Algorithm & Dominant term & Flops ($K_{\mathrm{res}}\!=\!8$) \\
        \midrule
        TSDCE-TD (Dict) &
            $\mathcal{O}\!\left(DQP\log QP + K_\mathrm{ref}\,D(n_{\mathrm{r}} n_{\mathrm{t}} + G_d)\right)$ &
            $\approx 9.4\!\times\!10^5$ \\[3pt]
        TSDCE-TD (FFT) &
            $\mathcal{O}\!\left(DQP\log QP + K_\mathrm{ref}\,D(n_{\mathrm{r}} n_{\mathrm{t}} + \log D)\right)$ &
            $\approx 8.6\!\times\!10^5$ \\[3pt]
        OMP &
            $\mathcal{O}\!\left(G_p^2\,G_d\cdot QP\cdot D\right)$ &
            $\approx 2.7\!\times\!10^{12}$ \\[3pt]
        MOMP &
            $\mathcal{O}\!\left(G_{p,\mathrm{lr}}^2 G_{d,\mathrm{lr}}(QP+D)
            + N_\mathrm{ref}(2G_p+G_d)QP\right)$ &
            $\approx 4.6\!\times\!10^9$ \\
        \bottomrule
    \end{tabular}}
\end{table}

\section{Conclusion}
\label{sec:conclusion}

In this paper, we proposed TSDCE-TD, a wideband time-domain extension of the Transformed Spatial Domain Channel Estimation~(TSDCE) algorithm for sparse channel estimation in millimeter-wave~(mmWave) integrated sensing and communication~(ISAC) systems.
Two variants were presented: TSDCE-TD~(Dict), which uses a one-dimensional delay dictionary, and TSDCE-TD~(FFT), which replaces the dictionary with Fast Fourier Transform~(FFT) zero-padding interpolation.
Both variants decouple angle and delay estimation through a matched-filter collapse in the beam-delay domain combined with successive interference cancellation~(SIC), yielding complexities several orders of magnitude lower than orthogonal matching pursuit~(OMP) and Multidimensional OMP~(MOMP) baselines at equivalent grid resolution.
Simulation results demonstrated that the proposed methods outperform the baselines in terms of channel normalized mean squared error~(NMSE) and angle root mean squared error~(RMSE)
under standard dictionary sizes, while achieving comparable performance when the baselines operate with larger dictionaries.
The methods were further validated in a ray-traced indoor scenario, where the estimated path parameters enabled accurate single-anchor user localization.
Future work will consider more realistic asynchronous scenarios and multipath-aided localization, as well as improved robustness in the low signal-to-noise ratio~(SNR) range.

\bibliographystyle{IEEEtran}
\bibliography{references}

\begin{thebibliography}{10}
\providecommand{\url}[1]{#1}
\csname url@samestyle\endcsname
\providecommand{\newblock}{\relax}
\providecommand{\bibinfo}[2]{#2}
\providecommand{\BIBentrySTDinterwordspacing}{\spaceskip=0pt\relax}
\providecommand{\BIBentryALTinterwordstretchfactor}{4}
\providecommand{\BIBentryALTinterwordspacing}{\spaceskip=\fontdimen2\font plus
\BIBentryALTinterwordstretchfactor\fontdimen3\font minus
  \fontdimen4\font\relax}
\providecommand{\BIBforeignlanguage}[2]{{%
\expandafter\ifx\csname l@#1\endcsname\relax
\typeout{** WARNING: IEEEtran.bst: No hyphenation pattern has been}%
\typeout{** loaded for the language `#1'. Using the pattern for}%
\typeout{** the default language instead.}%
\else
\language=\csname l@#1\endcsname
\fi
#2}}
\providecommand{\BIBdecl}{\relax}
\BIBdecl

\bibitem{hong2021role}
I.~A. Hemadeh, K.~Satyanarayana, M.~El-Hajjar, and L.~Hanzo, ``Millimeter-wave
  communications: Physical channel models, design considerations, antenna
  constructions, and link-budget,'' \emph{IEEE Communications Surveys \&
  Tutorials}, vol.~20, no.~2, pp. 870--913, 2018.

\bibitem{Gon24}
N.~Gonz{\'a}lez-Prelcic, M.~F. Keskin, O.~Kaltiokallio, M.~Valkama, D.~Dardari,
  X.~Shen, Y.~Shen, M.~Bayraktar, and H.~Wymeersch, ``{The integrated sensing
  and communication revolution for 6G: Vision, techniques, and applications},''
  \emph{Proceedings of the IEEE}, vol. 112, no.~7, pp. 676--723, 2024.

\bibitem{liu2025sensing}
F.~Liu, Y.-F. Liu, Y.~Cui, C.~Masouros, J.~Xu, T.~Xiao~Han, S.~Buzzi, Y.~C.
  Eldar, and S.~Jin, ``Sensing with communication signals: From information
  theory to signal processing,'' \emph{IEEE Journal on Selected Areas in
  Communications}, vol.~44, pp. 1--30, 2026.

\bibitem{momp}
J.~Palacios, N.~Gonz{\'a}lez-Prelcic, and C.~Rusu, ``Low complexity joint
  position and channel estimation at millimeter wave based on multidimensional
  orthogonal matching pursuit,'' in \emph{2022 30th European Signal Processing
  Conference (EUSIPCO)}, 2022, pp. 1002--1006.

\bibitem{nokia2026}
A.~Felix, R.~Hoffmann, M.~Henninger, S.~Ten~Brink, and S.~Mandelli, ``Angular
  estimation comparison with {ISAC} {PoC},'' in \emph{2026 IEEE 6th
  International Symposium on Joint Communications \& Sensing (JC\&S)}, 2026,
  pp. 1--6.

\bibitem{Barbieri_JC&S2026}
L.~Barbieri, M.~Henninger, P.~Tosi, A.~Grudnitsky, M.~Brambilla, M.~Nicoli, and
  S.~Mandelli, ``Target classification for integrated sensing and communication
  in industrial deployments,'' in \emph{2026 IEEE 6th International Symposium
  on Joint Communications \& Sensing (JC\&S)}, 2026, pp. 1--6.

\bibitem{tong2026distributed}
X.~Tong, Z.~Zhang, Z.~Yang, Y.~Ge, and H.~Wymeersch, ``Distributed multi-view
  environment sensing in wireless communication networks,'' \emph{IEEE
  Transactions on Wireless Communications}, vol.~25, pp. 12\,016--12\,033,
  2026.

\bibitem{tsdce}
S.~Roger, M.~Cobos, C.~Botella-Mascarell, and G.~Fodor, ``Fast channel
  estimation in the transformed spatial domain for analog millimeter wave
  systems,'' \emph{IEEE Transactions on Wireless Communications}, vol.~20,
  no.~9, pp. 5926--5941, 2021.

\bibitem{venugopal2017channel}
K.~Venugopal, A.~Alkhateeb, N.~G. Prelcic, and R.~W. Heath, ``Channel
  estimation for hybrid architecture-based wideband millimeter wave systems,''
  \emph{IEEE Journal on Selected Areas in Communications}, vol.~35, no.~9, pp.
  1996--2009, 2017.

\bibitem{COMST2025}
L.~Italiano, B.~{Camajori Tedeschini}, M.~Brambilla, H.~Huang, M.~Nicoli, and
  H.~Wymeersch, ``A tutorial on {5G} positioning,'' \emph{IEEE Communications
  Surveys \& Tutorials}, vol.~27, no.~3, pp. 1488--1535, 2025.

\bibitem{Zhang2016}
C.~Zhang, D.~Guo, and P.~Fan, ``Tracking angles of departure and arrival in a
  mobile millimeter wave channel,'' in \emph{2016 IEEE International Conference
  on Communications (ICC)}, 2016, pp. 1--6.

\bibitem{Shahmansoori2018}
A.~Shahmansoori, G.~E. Garcia, G.~Destino, G.~Seco-Granados, and H.~Wymeersch,
  ``Position and orientation estimation through millimeter-wave {MIMO} in {5G}
  systems,'' \emph{IEEE Transactions on Wireless Communications}, vol.~17,
  no.~3, pp. 1822--1835, 2018.

\bibitem{ETSI22261}
\BIBentryALTinterwordspacing
{ETSI}, ``{5G; Service Requirements for the 5G System},'' {European
  Telecommunications Standards Institute (ETSI)}, TS 122 261 v17.15.0, 2025.
  [Online]. Available:
  \url{https://www.etsi.org/deliver/etsi_ts/122200_122299/122261/17.15.00_60/ts_122261v171500p.pdf}
\BIBentrySTDinterwordspacing

\bibitem{SUN2024104372}
T.~Sun, A.~Peng, and J.~Shi, ``Multipath-assisted positioning enhancement via
  {DC-SLAM} with {MAP-PF}: An in-depth analysis and performance evaluation,''
  \emph{Digital Signal Processing}, vol. 146, p. 104372, 2024.

\end{thebibliography}

\end{document}